\documentclass[showpacs,twocolumn,preprintnumbers,amsmath,amssymb]{revtex4}
\usepackage{epsfig}
\usepackage{graphics}
\usepackage{color}

\newcommand{\etal}[1]{{\it et al.}}

\newcommand{\NDDD}{ND$_3$}

\begin{document}

\title{Continuous loading of an electrostatic trap for polar molecules}
\author{T. Rieger}
\author{T. Junglen}
\author{S.A. Rangwala}
\altaffiliation[New address: ]{Raman Research Institute, C. V.
Raman Avenue, Sadashivanagar, Bangalore 560080, India}
\author{P.W.H. Pinkse}
\author{G. Rempe}
\affiliation{Max-Planck-Institut f\"ur Quantenoptik,
Hans-Kopfermann-Str. 1, D-85748 Garching, Germany}
\date{\today, PREPRINT}

\begin{abstract}

A continuously operated electrostatic trap for polar molecules is
demonstrated. The trap has a volume of $\approx$ 0.6\,cm$^3$ and
holds molecules with a positive Stark shift. With deuterated
ammonia from a quadrupole velocity filter, a trap density of
$\approx10^8$\,cm$^{-3}$ is achieved with an average lifetime of
130\,ms and a motional temperature of $\approx 300$\,mK. The trap
offers good starting conditions for high-precision measurements,
and can be used as a first stage in cooling schemes for molecules
and as a ``reaction vessel'' in cold chemistry.


\end{abstract}

\maketitle

\vspace{1cm}

The production and storage of cold polar molecules is of
considerable interest to physicists and chemists. Once
sufficiently cold and dense samples are available, the anisotropic
and long-range dipole-dipole interaction can lead to new phases of
matter~\cite{Santos00,Baranov02,Goral02}. The interaction can even
be tuned by external fields allowing one to link polar
molecules~\cite{Bohn03}. Other applications include the study of
ultracold chemical reactions~\cite{Balakrishnan01}. Different
methods have been developed to produce and trap cold polar
molecules~\cite{EPJD04}. Here, trapping is defined as the ability
to store the particles much longer than the time it would take
them to leave the trap volume in the absence of the trapping
potential. So far, magnetic trapping was demonstrated for
molecules cooled by a buffer-gas~\cite{Weinstein98} or synthesized
from alkali atoms by photoassociation~\cite{Wang04}. In other
experiments decelerated samples of cold molecules were trapped in
static~\cite{Meijer-trap00,BethlemPRA02} or oscillating electric
fields~\cite{Veldhoven05}. A basic feature of all these methods is
their pulsed operation, where the trap is switched on after the
sample has been produced, or the sample is produced in the trap
center. From that point onward, the sample decays, e.g. by
collisions with background gas, and has to be regenerated.
Accumulative methods have been proposed, but remain
challenging~\cite{vandeMeerakkerPRA01}. Such continuously operated
traps would be advantageous not only for high-precision
measurements where long observation times with samples under
constant conditions are required, but also for determining cold
collision rates of reacting molecular species.

In this Letter we report on a novel electrostatic trap for polar
molecules, which is continuously loaded from an electrostatic
quadrupole guide~\cite{Rangwala03}. Equilibriation of the filling
and loss rates results in a steady state population of trapped
molecules. The trap works along electrostatic principles already
proposed some time ago ~\cite{Wing80}. It confines
low-field-seeking molecules in a region with low electric field
strength, surrounded by a region with a high electric field. The
trap is experimentally demonstrated with deuterated ammonia
(\NDDD), but can be used for all molecules exhibiting significant
Stark effect.
%
\begin{figure}[tb]\begin{center}
\epsfig{file=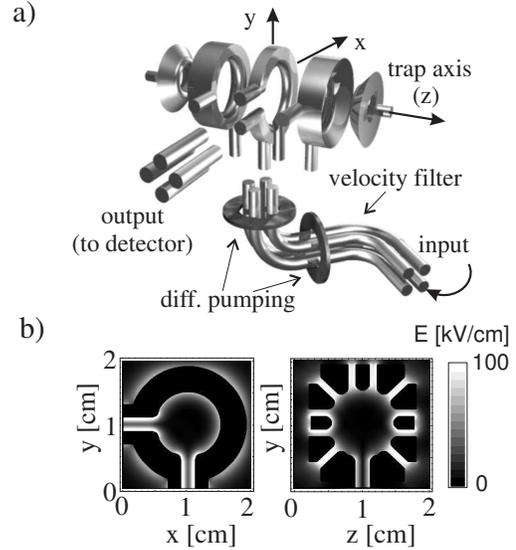,width=0.37 \textwidth}%
\caption{(a) Exploded view of the trap with input and output
quadrupoles. The quadrupole guide filters slow molecules from the
cooled effusive source and guides them into the trap. Here the
slowest molecules reside till they find their way out through one
of the two quadrupole guides. The trapped gas is analyzed with a
mass spectrometer positioned behind the output quadrupole. (b)
Electric field distribution in the xy plane for electrode voltages
of $\pm$5\,kV (left) together with the field distribution in the
yz plane containing the input quadrupolar guide (right).
}%
\label{trap}
\end{center}\end{figure}
%
As shown in Fig.~\ref{trap} a), the trap consists of five
ring-shaped electrodes and two spherical electrodes at both ends.
Neighboring electrodes carry high voltages of different polarity
giving rise to an inhomogeneous electric field, illustrated in
Fig.~\ref{trap} b), which is large near the electrodes and small
in the center of the trap. The central ring electrode is
intersected two times and local thickening of the neighboring
electrodes towards the gaps makes it possible to adapt two
quadrupole segments, one for filling and one for extraction of the
trapped gas. The inner radii of the five ring electrodes are (2.0;
4.3; 4.8; 4.3; 2.0)\,mm and the two end electrodes have a diameter
of 2\,mm. The electrodes are separated by a 1\,mm gap and the
enclosed volume amounts to $\approx$0.6\,cm$^3$. With the present
setup, a voltage difference of 10\,kV between neighboring trap
electrodes can be reached. This results in a minimum electric
field of $\approx$40\,kV/cm which the molecules need to overcome
in order to escape the trap in regions away from the small
entrance and exit holes. At these electric fields, only guided
molecules with velocities below $\approx$30\,m/s can be kept
within the trap where the maximum capture velocity depends on the
Stark shift of the individual molecule.

Once inside the trap, those molecules which overcome the Stark
potential barrier of the trap are either lost by hitting the
electrodes or they escape the trap and are pumped away. The
remaining molecules undergo multiple reflections off the trap
potential, randomizing their motion, and escape via the entrance
or exit hole. As the trap surface is large compared to the exit
channel area given by the two quadrupolar openings, the
probability of finding these holes is small and can even be made
arbitrary small by making the trap volume and/or the electric
field larger. Even though the trap allows continuous filling,
continuous accumulation is not possible in this conservative
potential and the trap density equilibrates when the filling rate
equals the leak-out rate.

In our experiment, the trap is filled with ND$_3$ from a
quadrupole velocity filter as detailed in \cite{Junglen04}. In
brief, molecules from a thermal reservoir are loaded into a
quadrupolar electric field guide via an appropriate nozzle
assembly. The quadrupole potential provides filtering of the
slowest molecules in the transverse direction whereas the
longitudinal filtering is achieved by the centripetal action in a
bent part of the guide between the nozzle and the trap. With this
technique, a continuously guided flux of the order of
10$^{10}$\,s$^{-1}$ can be achieved for quadrupole voltages of
$\pm$5\,kV, which results in a maximum electric field of
$\approx$90\,kV/cm. The guided flux consists of a mixture of
states with different Stark shifts with the larger Stark shifted
states preferentially guided. The longitudinal velocity
distribution can be described by a one-dimensional thermal
distribution with a most probable velocity of $\approx$40\,m/s.
Due to the two-dimensional confinement at finite field strengths,
the transversal velocity distribution is expected to be much
narrower. In passing we note that the Stark state distribution in
the guide and the trap is the same despite additional filtering in
the trap as discussed above.

As the average electric field inside the quadrupole is higher than
inside the trap, the molecules are accelerated when entering the
trap. It follows that the lowest-velocity molecules are absent in
the trap. The properties of the trapped sample are revealed by the
molecules which leave the trap through the 17\,cm long output
quadrupole. Here, the molecules are guided through a differential
pumping aperture into a separate vacuum chamber where they are
detected by a mass spectrometer (MS). Both the input and output
quadrupole guides are separated by a 0.5\,mm gap from a short
piece of a quadrupole guide formed from the trap electrodes. This
separation allows independent switching of the quadrupole segments
and the trap. The background pressure in the trap chamber is of
the order of 10$^{-10}$\,mbar and even lower in the detection
chamber.

\begin{figure}[htb]\begin{center}
\epsfig{file=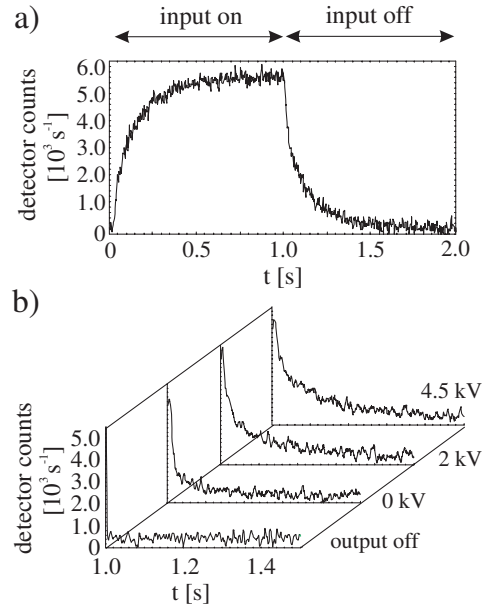,width=0.35 \textwidth}%
\caption{(a) The raw detector signal at the output quadrupole as a
function of time when the voltage at the input quadrupole is
switched. The electrodes carry voltages of $\pm$4.5\,kV. (b)
Deconvoluted detector signals as a function of time for different
trapping-field configurations. A rapid decay is observed when the
output is switched off at t$=$1\,s. The trap dynamics is revealed
by rapidly switching the voltage on one small ring electrode from
4.5\,kV to 0, 2 and 4.5\,kV, respectively, after finishing the
filling process in order to create an artificial hole in the trap.
When the particular electrode is set to 0\,kV a clear signal loss
can be observed, whereas the slow decay caused by trapped
molecules becomes dominant for higher voltages when the trap hole
is closed. The signal trace obtained with output off serves as a
reference.
}%
\label{decay}
\end{center}\end{figure}
In a first trapping experiment, voltages of $\pm$4.5\,kV were
applied to the trap and the output electrodes while the input
quadrupole was switched from 0\,kV to $\pm$4.5\,kV and back every
2\,s. The effusive source temperature was set to a constant value
of 160\,K. Fig~\ref{decay}(a) shows a raw trapping signal obtained
after averaging over 4000 cycles.  When the input quadrupole is
switched on at t$=$0, slow molecules are guided and the trap is
filled resulting in an increasing signal at the output quadrupole
guide. After the input quadrupole is switched off at t$=$1\,s, a
signal decay is observed which allows to estimate the lifetime of
the molecules in the trap. However, we found that the rising and
the falling slope show significant rate of change in signal
 even 500\,ms after switching on or off the input
quadrupole. After 500\,ms of filling or emptying of the trap,
steady state conditions are expected to prevail. Hence, a change
of the output flux cannot be caused by the molecules of interest.
On the rising slope, the excess signal is likely to be caused by a
local pressure increase near the detector after switching on the
guiding process. These molecules must be pumped away after
switching off the input guide, an effect which leads to a
time-dependent signal at the falling slope. These spurious signals
can be measured by switching on and off the output quadrupole
while the trap is continuously filled: As soon as the output guide
is switched off, a rapid decay of the MS signal to the background
level is expected, only limited by the time the molecules that
already left the guide need to fly to the detector, typically less
than 1.0\,ms. Indeed, a fast decay can be observed but it is
accompanied by a slow decay with a 1/e-time of $\approx$150\,ms.
Assuming that the guided flux into the detector is a sharp step
function $\Theta(t)$, the measured signal $S_{\rm \Theta}(t)$ is
used to determine the deconvolution kernel, ${\rm F}(K)={\rm
F}(S_{\rm \Theta}(t))/{\rm F}(\Theta(t))$, where ${\rm F}$ denotes
the Fourier transform. For determining the lifetime of the
molecules in the trap, the measured decay signal, $S_{\rm D}(t)$,
is deconvoluted by the transformation $S(t)={\rm F^{-1}}[{\rm
F}(S_{\rm D}(t))/{\rm F}(K)]$, where ${\rm F^{-1}}$ denotes the
inverse Fourier transform. After deconvolution the slow rise and
fall of the signal $500\,$ms after switching on and off the input
guide vanishes.

Having described the deconvolution process, we now discuss the
trap measurements in which the input quadrupole is periodically
switched. In order to demonstrate that the decay rate originates
from the trap dynamics, an artificial hole was created in the trap
by rapidly lowering the voltage on one small ring electrode to a
constant value when the input quadrupole and accordingly the
filling process is switched off. Fig.~\ref{decay}(b) shows the
decay signals after deconvolution for reduced voltages on the
particular ring electrode of 0, 2 and 4.5\,kV, respectively. In
the decay measurement where the voltage has been switched to
0\,kV, a fast decay can be observed followed by a small and slow
decay contribution. The fast decay is caused by the losses due to
the weaker field near the particular ring electrode. The molecules
causing the slow decay are either too slow to overcome even the
weak field potential barrier or they do not encounter the weak
field on their way inside the trap. Note that the start of the
decay is delayed by the time the molecules need to pass through
the output quadrupole. When the voltage on the ring electrode is
raised to 2.0\,kV, the loss rate decreases which leads to a
reduction of the fast decay contribution whereas the slow decay
caused by molecules which are trapped longer is more pronounced.
For the remaining curve the voltage on the electrode is set to
4.5\,kV and here the slow decay is dominant.

Note that the initial fast decay can always be observed.
Simulations show that this decay is caused by a class of molecular
trajectories which approximately are confined in the plane defined
by the middle ring electrode. As both exit channels lie in this
plane a fast escape is very probable. For those molecules whose
trajectories fill the whole trap volume the escape probability is
reduced leading to a longer trapping time. As the lifetime of the
molecules in the trap depends on how fast they find an exit, the
lifetime is velocity dependent. Therefore, the decay cannot be
described by an exponential function and, hence, not by a
(1/e)-lifetime. An alternative measure for the trap lifetime is
the time after which half the molecules have left the trap. From
the data for $\pm 4.5\,$kV a lifetime of 130 $\pm$10\,ms can be
derived.

The lifetime is mainly limited by the exit channels whereas
collisions with the background gas do not contribute
significantly. As the field vanishes only at some regions in the
center of the trap, Majorana transitions to nontrapping Stark
levels are not very likely. With the angle distribution of the
guided molecules behind the output quadrupole and the sensitivity
of the MS the total flux emerging from the trap can be determined.
From the measured angle distribution it has been determined that
only 15 percent of the guided flux reaches the detector. As the
detector sensitivity is of the order of 10$^{-4}$\,counts/molecule
the guided flux from the trap with all electrodes set to
$\pm$4.5\,kV amounts to 3$\times$10$^8$\,s$^{-1}$. Similar
trapping results were obtained with formaldehyde (CH$_2$O) and
methylchloride (CH$_3$Cl) which also show a linear Stark effect.

%
\begin{figure}[htb]\begin{center}
\epsfig{file=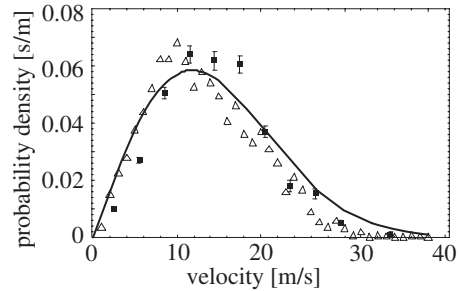,width=0.33 \textwidth}%
\caption{Velocity distribution of trapped ND$_3$ molecules derived
from data obtained with electrode voltages of $\pm$\,4.5\,kV
(squares). The line is a fit to the measured data of the
functional form $(2 v/\alpha^2) \exp[-v^2/\alpha^2]$, with the
characteristic longitudinal velocity $\alpha=16.6\pm 1\,$m/s. The
triangles denote simulation results.} \label{velocity}
\end{center}\end{figure}
%
In the following the temperature of the trapped ND$_3$ sample is
determined by a measurement of the molecules' velocity
distribution. Therefore, a time-of-flight measurement is performed
where the trap is continuously filled and only the output
quadrupole is switched on and off at rate 1\,Hz. All electrodes
were set to voltages of $\pm$4.5\,kV. As soon as the output
quadrupole is switched on, molecules from the trap are guided to
the detector where their arrival time is recorded. After more than
50,000 cycles a clear time-of-flight signal has developed. From
the delay and the rising slope the longitudinal velocity
distribution can be derived by differentiation. For a single
molecular state, the velocity distribution should show a
relatively sharp velocity cut-off because the guide and the trap
filter on kinetic energy. But given a mixture of states with
different Stark shifts, the cut-off is smeared out so that the
velocity distribution can be described by a one-dimensional
thermal distribution. As the time-of-flight signal is affected by
the above-mentioned spurious signal, the measured signal was
deconvoluted according to the method described above.
Fig.~\ref{velocity} shows the velocity distribution obtained from
the deconvoluted data. It can be described by a characteristic
velocity $\alpha=16\pm 1$\,m/s with $\alpha$=$\sqrt{2 k_B T/m}$,
where $k_B$ is the Boltzmann constant, $T$ the temperature and $m$
the molecular mass. The characteristic velocity $\alpha$ is
equivalent to the most probable speed of a thermal gas in a
three-dimensional volume element. This velocity corresponds to a
motional temperature of 300\,mK. Note that the lowest velocity
molecules may be partially depleted by collisions in the nozzle
where the densities are relatively high. Note also that the
molecular velocities in the output quadrupole are slightly smaller
than inside the trap because the molecules are decelerated when
entering the higher quadrupole field. However, as there are only
conservative potentials involved, the temperature of the sample
does not change. The experimental data are in good agreement with
the simulation. As expected, the temperature of 300\,mK is smaller
than the trap depth of 800\,mK derived from the average Stark
potential.

The transverse velocity distribution of the trapped molecules can
be estimated by recording the decrease of the flux of guided
molecules in the output quadrupole when the electric field in the
latter is reduced. During the measurements, the voltages on the
input quadrupole and the trap were set to $\pm$4.5\,kV and the
input was modulated. In each measurement the electric field in the
output quadrupole was set to a different but constant value. It
was observed that more than 90$\%$ of the flux from the trap can
be guided even if the output quadrupole voltage is reduced down to
$\pm$\,750\,V so that most of the molecules can be
two-dimensionally trapped in an electric field of only 15\,kV/cm.
Below this voltage, the guided flux decreases and at voltages of
$\pm$160\,V ($\approx$3\,kV/cm), for example, the initial flux has
reduced by a factor of two. The measured decrease of the signal
amplitude as a function of the output quadrupole voltage is in
good agreement with the simulation of the experiment. This
justifies the assumption that the simulated transverse velocity
distribution with its characteristic velocity of
$\alpha_{sim}=14\,$m/s describes the experiment well. It follows
that the characteristic velocity $\alpha$ of the longitudinal
distribution is roughly equal to the characteristic velocity
$\alpha_{sim}$ determined for the transversal velocity
distribution, which is in contrast to the situation in the input
quadrupole guide. There, the longitudinal velocities can be much
higher as longitudinal filtering is less restrictive than
transverse filtering. The trap randomises the input velocity
which leads to an equilibration of the transverse and longitudinal
velocity distributions, as is also obtained in the simulation.
With the knowledge of the average speed
$\bar{v}=2\alpha/\sqrt{\pi}$ and the total molecular flux $\Phi$
out of the trap, one can estimate the number density $n$ in the
trap by $\Phi=\frac{1}{4}n \bar{v} A$ where $A$ is the effective
area of each of the two exit channels. From simulations we know
that the molecular density distribution inside the guide with
voltages of $\pm$4.5\,kV has a half width of $\approx$400\,$\mu$m.
Taking this as the radius of a circular area A and assuming an
average velocity $\bar{v}$ of 18\,m/s, the number density inside
the trap is of the order of 10$^8$\,cm$^{-3}$.

To summarize, a continuously loaded, large-volume electrostatic
trap for polar molecules has been demonstrated experimentally.
Our results show that a sample of ND$_3$ molecules at a density
of 10$^8$\,cm$^{-3}$ can be trapped with a lifetime of 130
$\pm$10\,ms. The trap is filled from a quadrupole guide, but it
is also conceivable to produce molecules inside the trap, for
example by reactive collisions~\cite{Loesch} or crossed
molecular beams~\cite{Elioff03}. These molecules could be cooled
by a buffer gas~\cite{Weinstein98} or light
scattering~\cite{Vuletic00}. A simple modification of the
present trap with three guides coupled to it could lead to the
intriguing possibility of an electrostatic ``reaction vessel''.
Two of the guides could be used to load different species of
molecules into the trap and the third guide could be used to
extract and detect the reaction products. If the trap volume is
large, the lifetime and the number of molecules is expected to
increase so that reactions inside the trap become possible even
at the present densities. Provided electrostatic or mechanical
valves can be developed, one can even envision a small cold
chemical factory made out of a network of interconnected
reaction vessels with new possibilities in controlling chemical
reactions.


\end{document}